\documentstyle[floats,prb,aps]{revtex}

\input epsf
\begin{document}
\draft
\twocolumn[\hsize\textwidth\columnwidth\hsize\csname @twocolumnfalse\endcsname
\title{Exact Solution of the strong coupling t-V model \\
with twisted boundary conditions}
\author{R. G. Dias}
\address{Departamento de F\'{\i}sica, Universidade de Aveiro, \\
3810 Aveiro, Portugal.}
\date{\today}
\maketitle
\widetext
\begin{abstract}
\leftskip 54.8pt \rightskip 54.8pt We present the solution of the
one-dimensional t-V model with twisted boundary conditions in the
strong coupling limit, $t \ll V$ and show that  this model can be
mapped onto the strong coupling Hubbard chain threaded by a
fictitious flux proportional to the total momentum of the charge
carriers. The high energy eigenstates are characterized by a
factorization of degrees of freedom associated with configurations
of soliton and antisoliton domains and degrees of freedom
associated with the movement of ``holes'' through these domains.
The coexistence of solitons and antisolitons leads to a
strange flux dependence of the eigenvalues. We illustrate the use
of this solution, deriving the full frequency dependence of the
optical conductivity at half-filling and zero temperature.

\end{abstract}

\pacs{\leftskip 54.8pt
\rightskip 54.8pt
PACS numbers: }
]
\narrowtext

\section{Introduction}
The extended Hubbard model and its spinless version, the t-V model,
have been extensively studied due to their relevance in the comprehension
of the behavior
of strongly correlated compounds such as cuprates\cite{ande1,ande2,zhang}
and organic conductors. \cite{dardel,schulz}
Much of the present understanding of these models has been a
consequence of the exact solution
in one dimension by the Bethe ansatz (BA) technique.
\cite{bethe,lieb}
The evaluation of the correlations
remains however a hard task within the Bethe ansatz framework.
For the Hubbard model, further progress was possible in the
strong coupling limit due to the simpler form of the solution.
\cite{ogat,penc1,penc2,eske,paro}
The eigenfunctions in this limit factorize as a product of a
wavefunction of non-interacting spinless fermions and a wavefunction
of a squeezed spin chain.\cite{ogat,dias1}
This spin-charge factorization simplifies the calculation of correlations
and in particular, it has been used to determine
the momentum distribution function,\cite{ogat}
the spectral function,\cite{penc1,penc2}
the sum rules of the upper and lower Hubbard bands \cite{eske} and
the Green's function \cite{paro} of this model.
An alternative solution to that of the Bethe ansatz was also possible
in this limit.\cite{dias1,nuno1,gebh}

The t-V model is apparently simpler than the Hubbard model due to
the absence of spin degrees of freedom. This model can be mapped
onto the anisotropic Heisenberg model (more precisely, the XXZ or
Heisenberg-Ising model) by the Jordan-Wigner
transformation,\cite{jord} whose Bethe ansatz solution has long
been known.\cite{xxz} In the strong coupling limit, the t-V model,
despite its apparent simplicity, remains somewhat foggier than the
Hubbard model. For instance, the Bethe ansatz
solution\cite{nuno2,suth,kusm} presents us eigenvalues expressions
with phase terms whose physical meaning is not clear. Another
curious fact is that the Luttinger liquid exponent\cite{hald} that
characterizes the low energy excitations of the strong coupling
t-V model is density dependent \cite{gomez} in contrast to the
strong coupling Hubbard model where it is a constant. \cite{ande2}
Since this exponent is closely related to Fermi surface phase
shifts (a holon Fermi surface in the case of the Hubbard model),\cite{ande2}
it is worthwhile to investigate how these phase shifts will be modified.
In this paper, we present a non-Bethe-ansatz solution for the
strong coupling one-dimensional t-V model which is closely related
to the solution of the strong coupling Hubbard model \cite{nuno1}
and which clarifies the previous issues. The simple factorized
form of this solution (and the low degeneracy of the eigenvalues)
will, we believe, allow an easy calculation of correlations.

The  t-V Hamiltonian for a ring with
$L$ sites is
\begin{equation}
     H = -t \sum_{i} (c_{i}^\dagger c_{i+1}
     + c_{i+1}^\dagger c_{i})
     + V \sum_i n_{i} n_{i+1},
     \label{eq:H1}
\end{equation}
where $c_{i}^\dagger$ is the fermion creation operator on site
$i$, $n_{i}=c_{i}^\dagger c_{i}$
and $V$ is the nearest-neighbor Coulomb interaction.
The  one-dimensional strong coupling t-V  model (as the Hubbard model)
is a classic example of a system which exhibits a metal-insulator transition
upon doping.
If $t=0$, the fermions are localized  and all states 
with the same number of 
pairs of nearest-neighbor occupied sites, $\sum_i n_{i} n_{i+1}$, 
are degenerate. This degeneracy is lifted if $t/V$ is finite and
up to first order in $t$, the eigenvalues are obtained diagonalizing the 
Hamiltonian within each of the degenerate subspaces.
In the strong coupling limit $t \ll V$, we obtain therefore 
the projected Hamiltonian
\begin{eqnarray}
     H &=& -t \sum_{i} [(1-n_{i+2}) c_{i+1}^\dagger c_{i} (1-n_{i-1})
     + hc] \nonumber \\
     && -t \sum_{i} [n_{i+2} c_{i+1}^\dagger c_{i} n_{i-1}
     + hc] \nonumber \\
     && +V \sum_i n_{i} n_{i+1}.
\label{eq:H2}
\end{eqnarray}
This limit corresponds to the 
$J_x=J_y \ll J_z$ limit of the anisotropic Heisenberg model.
The set of eigenstates and
eigenvalues of this model can be obtained without having to resort to
the Bethe ansatz,
as we will show below.

The behavior of the t-V model in the strong coupling limit has
provided support for a recent conjecture by Zotos and Prelovsek.
\cite{zotos,cast,naef} According to these authors, the t-V model
and the Hubbard model at half-filling are perfect insulators, this
meaning that the Drude weight ($D_c$) in the thermodynamic limit
remains zero even at finite temperature. In particular, in the
strong coupling limit, they argue that $D_c=0$ even for finite
systems. Such behavior was confirmed by Peres {\it et al},
\cite{nuno2} applying the Bethe ansatz method to solve the t-V
model in the strong coupling limit. Here, we present a different
solution which allows an easy confirmation of the previous results
and makes clearer the physical picture in this limit.

The remaining part of this paper is organized in the following
way. In section II, the low energy eigenstates of the model are
found. First, we study the one-particle problem and present a new
path for its solution. Then, we show  that this solution can be
extended to the case with $N$ particles. We also find the flux
dependence of the eigenvalues. In section III, the general
solution is presented both for periodic and twisted boundary
conditions (finite flux). We also comment on the higher order
corrections. In section IV, we compare our results with those
obtained with the Bethe ansatz technique. The transport properties
of the model are studied in section V. Finally, we conclude in
section VI.

\section{low energy subspace}
Let us consider two consecutive sites and therefore, nearest neighbors of
each other. There are four different configurations for this pair of sites,
which we will call links and they are
\begin{equation}
        (\bullet \, \bullet); (-  \, \bullet); (\bullet  \, -); (-  \, -)
\end{equation}
where a dot stands for an occupied site and a dash for an empty one.
The total number of these links in the chain is equal to the number of
sites $L$,
\begin{equation}
        N_{\bullet \, \bullet}+
        N_{-  \, \bullet}+N_{\bullet  \, -}+N_{-  \, -}=L
\end{equation}
and $N_{-  \, \bullet}=N_{\bullet  \, -}$. Further conditions result from
counting the number of holes or particles,
\begin{eqnarray}
        N_{-  \, \bullet}+N_{-  \, -} &=& N_h, \nonumber \\
        N_{\bullet \, \bullet}+N_{-  \, \bullet} &=& N_e .
\end{eqnarray}
In the limit $V/t \rightarrow \infty$, the number of
links $(\bullet \, \bullet)$
is a conserved quantity and consequently also are $N_{-  \, \bullet}$
and $ N_{-  \, -}$. So, in the strong coupling limit, the model 
merely exchanges the positions of these links.
Note that $\sum_i n_{i} n_{i+1}
=N_{\bullet \, \bullet}$.

\subsection{One particle}
Let us start with the simple case of a single particle in our periodic
chain. In this case, the interaction term is zero and we have a one-particle
tight-binding model, whose solution is trivial. We are going to solve this
model in a different fashion, considering as our mobile particle the
link $(-  \, \bullet)$.  This link moves exchanging its
position with a  link $(-  \, -)$.
Note that we have one link $(-  \, \bullet)$ and $L-2$ links $(-  \, -)$,
and therefore, the total number of these links is $\tilde{L}=L-1$.

First, let us define our states in terms of the position of this link,
\begin{equation}
        \vert \tilde{i} \rangle = c_{i+1}^\dagger \vert 0\rangle
\end{equation}
with $1 \leq \tilde{i} \leq L-1$. Note that there is a state
$c_{1}^\dagger \vert 0\rangle$ which is not included in the previous set
of states, but this state can be written as
\begin{equation}
        \hat{T}^{-1} \vert \tilde{1} \rangle =
        \hat{T}^{-1} c_{2}^\dagger \vert 0\rangle
\end{equation}
where $\hat{T}$ is the one-site translation operator.
The single particle Hamiltonian rewritten using this new notation becomes
\begin{equation}
     H/(-t) =
     \sum_{\tilde{i} \neq \tilde{L}} \vert \tilde{i}+1 \rangle \langle
     \tilde{i} \vert + \hat{T}^{-1} \vert \tilde{1} \rangle \langle
     \tilde{L} \vert +h.c.
\label{eq:H3}
\end{equation}
with $\tilde{L}=L-1$.
We now introduce an  over-complete set of states, constructing from a state
$\vert \tilde{i}\rangle$, a state invariant by translation with momentum
$k$,
\begin{equation}
        \vert \tilde{i},k \rangle = {1 \over \sqrt{L}} \sum_{j=1}^{L}
        e^{ikj} \hat{T}^{j-1} \vert \tilde{i}\rangle.
\end{equation}
These states diagonalize the Hamiltonian, but we are going to proceed as
if they did not and as if they were an orthogonal set of states.
The Hamiltonian becomes
\begin{eqnarray}
     H/(-t) & = &
     \sum_{k}
     \Big(
     \sum_{\tilde{i}  \neq \tilde{L}} \vert \tilde{i}+1,k  \rangle \langle
     \tilde{i},k  \vert  \nonumber \\
     && + e^{ik} \vert \tilde{1},k  \rangle \langle
     \tilde{L},k  \vert +h.c. \Big) \label{eq:H4}
\end{eqnarray}
The Hamiltonian in a given $k$ subspace has become that of
a tight-binding model in
a chain of $L-1$ sites with a fictitious flux $e^{ik}$. The eigenvalues
are given by
\begin{equation}
        E(\tilde{k},k)=-2t \cos \left(\tilde{k}-{k \over \tilde{L}} \right)
\end{equation}
with $\tilde{k}=\tilde{n} \cdot 2\pi /\tilde{L}$, 
$\tilde{n}=0,\dots \tilde{L}-1$
and $k=n \cdot 2\pi /L$, $n=0,\dots L-1$. But clearly, not all combinations
of $\tilde{k}$ and $k$ correspond to real eigenstates of the Hamiltonian.
The eigenstates will be of the form
\begin{equation}
        \vert \tilde{k},k \rangle = {1 \over \sqrt{\tilde{L}}} 
        \sum_{j=1}^{\tilde{L}}
        e^{i(\tilde{k}-{k \over \tilde{L}})j} \vert \tilde{j},k \rangle,
\end{equation}
but obviously, this is a combination of states which are the same state except
for a phase, that is,
\begin{equation}
       \vert \tilde{j},k \rangle = e^{ik}\vert \tilde{j}+1,k \rangle
\end{equation}
and the linear combination will be zero unless the following condition is
satisfied
\begin{equation}
        \tilde{k}-{k \over \tilde{L}} - k=0 \quad (\text{mod} \,\, 2\pi)
\end{equation}
which implies $\tilde{k}=k L/\tilde{L} \,\,(\text{mod} \,\, 2\pi)$.
This condition is equivalent
to stating that the final state must obviously have a momentum $k$.
Note that the previous equation has precisely $L$ solutions
and therefore, the usual set of tight-binding eigenvalues is recovered.

Let us make a few remarks concerning the above procedure. Let
$\{\vert i \rangle \}$, $i=1,\dots L$, be an orthogonal set of states
which constitutes a basis for the states of a given system and let $H_{ij}$
be the matrix elements of the system Hamiltonian between the states
$\vert i \rangle$ and $\vert j \rangle$. The set of eigenstates
$\{\vert \phi_i \rangle \}$, $i=1,\dots L$, of the Hamiltonian can be
written in this basis as $\vert \phi\rangle= \sum_j a_j \vert j \rangle$.
Assume now that two or more states of this basis were in fact the same
state. This would lead to a reduced matrix $\tilde{H}_{ij}$ which
would have the same elements, except for the transitions to and from the state
which remained from the set of equivalent states. These transitions
are multiplied by the total number of equivalent states.
The eigenstates of $H$ {\it remain} eigenstates of $\tilde{H}$, but now
they constitute an over-complete basis of the Hilbert space, that is,
they are not all linearly independent. They may however remain an orthogonal
set if the states in excess are identically zero as in the case studied above.

\subsection{several particles}
\label{sec:particles}
Let us consider now the case of $N_e$ particles in a chain of $L$ sites, but
distributed so that there are no links $(\bullet \, \bullet)$. These states
are of the form
\begin{equation}
        \vert a_1, \dots , a_{N_e}\rangle = \prod_{i=1}^{N_e}
        (1-n_{a_i-1}) c_{a_i}^\dagger \vert 0\rangle
\end{equation}
with $\{a_i\}$ an ordered set of non-consecutive numbers ($a_1 >1$).
The total number
of $(-  \, -)$ and $(-  \, \bullet)$ links is $\tilde{L}=L-N_e$.
This state can be mapped onto the following
state of a chain with $L-N_e$ sites,
\begin{equation}
        \vert \tilde{a}_1, \dots , \tilde{a}_{N_e}\rangle = \prod_{i=1}^{N_e}
        \tilde{c}_{\tilde{a}_i}^\dagger \vert \tilde{0}\rangle
\end{equation}
such that if the first site of this pseudo-chain is empty, the first
link of the $L$ sites chain is $(-  \, -)$, if it is occupied, the first
link is $(-  \, \bullet)$. The same reasoning applies to the other sites.
Note that as in the previous case, there are a few states which are not
included in the above set, namely states where  a link $(-  \, \bullet)$
is divided between sites $L$ and $1$. These states have a particle at
site $1$ and one should note that starting from a state as above, these
states appear when a link $(-  \, \bullet)$ is at site $\tilde{L}$,
or equivalently, a particle is  at site $L$ and hops to site $1$.
However, this hopping  term can written as
\begin{equation}
        \hat{T}^{-1} \tilde{c}_{\tilde{1}}^\dagger \tilde{c}_{\tilde{L}}
        \vert \tilde{a}_1, \dots , \tilde{a}_{N_e-1}, \tilde{L}\rangle
\end{equation}
where $\hat{T}$ is the one-site translation operator in the chain of $L$
sites. Note that this translation operator also insures that the other
pseudo-particles remain in the same sites in the reduced chain.
Given a state $\vert \tilde{a}_1, \dots , \tilde{a}_{N_e}\rangle$,
we build as previously the state invariant by translation with momentum $P$,
\begin{equation}
        \vert \tilde{a}_1, \dots , \tilde{a}_{N_e},P \rangle =
        {1 \over \sqrt{L}} \sum_{j=1}^{L}
        e^{iPj} \hat{T}^{j-1} \vert \tilde{a}_1, \dots , 
        \tilde{a}_{N_e}\rangle.
\end{equation}
The mapped Hamiltonian in the subspace of states with momentum $P$ becomes
\begin{equation}
     H(P) = -t \left( \sum_{\tilde{i}\neq \tilde{L}} 
     \tilde{c}_{\tilde{i}+1}^\dagger
     \tilde{c}_{\tilde{i}} + e^{iP}  \tilde{c}_{\tilde{1}}^\dagger
     \tilde{c}_{\tilde{L}} \right) +hc.
\end{equation}
So, we have mapped the Hamiltonian onto a tight-binding chain with $L-N$
sites threaded by a flux $P$ with $N_e$ particles.
The eigenvalues are given by
\begin{equation}
        E(\{\tilde{k}\},P)=-2t \sum_{i=1}^{N_e}
        \cos \left(\tilde{k}_i-{P \over \tilde{L}} \right)
        \label{eq:sevE}
\end{equation}
with $\tilde{k}=\tilde{n} \cdot 2\pi /\tilde{L}$,
and $P=n \cdot 2\pi /L$, with $\tilde{n}=0, \dots, \tilde{L}-1$
and $n=0, \dots, L-1$.
Again, not all combinations of pseudo-momenta
$\{\tilde{k}\}$ and $P$ are possible
and applying the same procedure as before, we arrive to following condition
\begin{equation}
        \sum_{i=1}^{N_e}(\tilde{k}_i-{P \over \tilde{L}})
         - P=0 \quad (\text{mod} \,\, 2\pi)
\end{equation}
which implies
\begin{equation}
        P {L \over \tilde{L}}=\sum_{i=1}^{N_e} \tilde{k}
        \quad (\text{mod} \,\, 2\pi).
\end{equation}
The factor $L/\tilde{L}$ converts the total momentum of our chain
of $L$ sites in the total momentum of the pseudo-chain. Note that
the set of pseudo-momenta $\{\tilde{k}\}$ is not enough to define
the total momentum $P$ since there may be two values of $P$ such that
$PL/\tilde{L}$ (mod $2\pi$) is the same. In fact, if $P=n \cdot 2 \pi/L$
,
with $n=0, \dots,L-1$, the states $\vert \{ \tilde{k}
\}, P \rangle$ with $n$ in the interval $[0,L-1-\tilde{L}]$ have
the same energy as the states $\vert \{ \tilde{k} \}, P+ {2\pi
\tilde{L} \over L} \rangle$. In the particular case of
$\tilde{L}=L/2$, given a state with momentum $P$, one always has a
state with momentum $P+\pi$ with the same set of wavenumbers. The
$\tilde{L}=L/2$ case corresponds to the half filling and indeed
one knows that the ground state is doubly degenerate, one state
having zero momentum, the other $\pi$.
This is also valid for 
excited states (with $N_{\bullet \, \bullet} \neq 0$), as we
shall see in the next section.

An external magnetic flux $\phi$ can be introduced in the problem with the
transformation $t\rightarrow te^{i\phi/L}$. The Hamiltonian remains invariant
by translation and all the previous steps can be repeated, leading to the
following modification in the eigenvalue expression
\begin{equation}
        E(\{\tilde{k}\}) \rightarrow E\left(\left\{\tilde{k}
        -{\phi \over L} \right\}\right)
.
\end{equation}

The ground state energy is given by Eq.~\ref{eq:sevE}. If $N_e$
is odd, all single-particle states with pseudo-momentum
$\tilde{k}$ between $\pm 2\pi/\tilde{L} \cdot (N_e-1)/2$ are
occupied and $\sum \tilde{k}=0$. Therefore,
\begin{eqnarray}
        E_{gs}^{odd} &=&
        -2t \sum_{i=-{N_e-1 \over 2}}^{{N_e-1 \over 2}}
        \cos \left({2 \pi \over \tilde{L}} i \right) \nonumber \\
        &=& -2t {\sin \left({\pi N_e \over \tilde{L}}\right)
        \over \sin \left({\pi \over \tilde{L}}\right)}
.
\end{eqnarray}
If $N_e$ is even, all states with
$\tilde{k}$ between $- 2\pi/\tilde{L} \cdot (N_e-2)/2$ and
$ 2\pi/\tilde{L} \cdot N_e/2$ or between $- 2\pi/\tilde{L} \cdot N_e/2$
and $ 2\pi/\tilde{L} \cdot (N_e-2)/2$
are occupied and $\sum \tilde{k}=\pm \pi/\tilde{L} \cdot N_e/L$. So,
\begin{eqnarray}
        E_{gs}^{even} &=& -2t \sum_{i=- {N_e-2 \over 2}}^{N_e/2}
        \cos \left({2 \pi \over \tilde{L}} i -{\pi \over \tilde{L}}
        {N_e \over L} \right) \nonumber \\
        &=& -2t {\sin \left({\pi N_e \over \tilde{L}}\right)
        \over \sin \left({\pi \over \tilde{L}}\right)}
        \cos \left({\pi \over L}\right)
.
\end{eqnarray}
This slight energy difference between the two cases had already
been pointed out by Kusmartsev. \cite{kusm} In the presence of a
small flux $\phi$, the $N_e$ odd expression should be multiplied
by a factor $\cos(\phi /L)$, while for $N_e$ even, a $-\phi /L$
term should be summed to the argument of the cosine.

This phase shift between the wavenumbers of the ground states with
$N$ and $N+1$ particles should be responsible for the
orthogonality catastrophe in the thermodynamic limit which, for example,
leads to a zero renormalization constant $Z$
characteristic of a Luttinger liquid \cite{hald}
(see Ref.~\cite{ande2} for a detailed
calculation in the case of the strong coupling Hubbard model).
The renormalization constant $Z$ is given 
by the overlap between the ground state
with $N+1$ particles and the ground state with $N$ particles plus a particle
at Fermi momentum,
\begin{equation}
     Z = \vert \langle\psi_{GS} (N+1;P=k_f)\vert
     c^{\dagger}_{k_F\sigma}\vert \psi_{GS} (N;P=0)\rangle\vert^2
\end{equation}
yielding zero in the thermodynamic limit. Our results above
indicate that the phase shift depends on the density, $\delta=\pi
{N_e \over L}= \pi \rho$. Recall that in the case of the strong
coupling Hubbard model, the phase shift of the holons wavenumbers
is independent of the band filling, $\delta={\pi \over 2}$. That
phase shift results from a $(-k_F)$ momentum contribution from the
spin sector.\cite{ande2} Here, the phase shift is due to the total
momentum of the charge carriers. This dependence on the band
filling is in agreement with the fact that the anomalous exponent
of this model is indeed band filling dependent. \cite{gomez} The
Luttinger liquid velocities\cite{hald} that characterize the low
lying excitations have been found by Gomez-Santos \cite{gomez} for
the  strong coupling t-V model in the thermodynamic limit based on very
simple arguments (basically, the reduction of the effective size
of the chain). These velocities and its finite size corrections
are easily obtained from the previous equations. In the large $L$
limit, the Gomez-Santos results are reproduced:
\begin{equation}
    v_N={L \over \pi} {\partial^2 E_T \over \partial N^2}
    ={2t \over (1- \rho)^3} \sin \left( { \rho\pi
    \over 1- \rho} \right)
\end{equation}
\begin{equation}
    v_J={L \over \pi} {\partial^2 E_T \over \partial \phi^2}
    =2t \left(1- \rho\right) \sin \left( { \rho\pi
    \over 1- \rho} \right)
\end{equation}
\begin{equation}
    v_S=\sqrt{v_N v_J}={2t \over (1- \rho)} \sin \left( { \rho\pi
    \over 1- \rho} \right)
\end{equation}
where $\rho={N_e \over L}$ and $v_N$, $v_J$, and $v_S$ are respectilvely
the particle, current, and sound-wave velocities.
\section{general solution}
Let us consider the  general case where one may have both links
$(\bullet \, \bullet)$ and $(-  \, -)$. First, note that a phase separated
state (one domain of  holes and one domain of particles) has no mobile
entities in the strong coupling limit since any hopping of a single particle
would imply the breakup of a link  $(\bullet \, \bullet)$.
So, phase separated states will be eigenstates of the 
strong coupling Hamiltonian with eigenvalues
given by $E=V\cdot N_{\bullet \, \bullet}$.
Furthermore, the same applies to states
with several domains if the only links $(- \, \bullet)$ present are the
domain walls. Clearly, a hole (particle),  in order to be able to move,
must be within a particle (hole) domain. If for a chain with $L$ sites
and $N_e$ particles, we fix $N_{\bullet \, \bullet}$ and $N_{-  \, -}$, it is
the configuration of these links that will define how many mobile
links $(- \, \bullet)$ one has and consequently, 
the number of sites $\tilde{L}$
of the effective chain for these mobile links. 
These mobile links will move exchanging
their position with links $(\bullet \, \bullet)$ and $(-  \, -)$.

It will prove itself useful to do the following mapping:
\begin{eqnarray}
        (\bullet \, \bullet)&=& \vert \uparrow \rangle , \nonumber \\
        (-  \, -) &=& \vert \downarrow \rangle , \nonumber \\
        (-  \, \bullet) &=& \vert 0 \rangle , \nonumber
\end{eqnarray}
with the exception of the links which are domain walls.
That is, we will map the states of the spinless chain with $L$ sites
and $N_e$ particles onto states of a spinful chain with 
$\tilde{L}$ sites and $N_{\bullet \, \bullet}$ particles with spin up
and $N_{-  \, -}$ particles with spin down.
The first two 
links are called respectively a soliton and an antisoliton.
A general state is written as
\begin{equation}
        \vert a_1, \dots , a_{N_e}\rangle = \prod_{i=1}^{N_e}
        c_{a_i}^\dagger \vert 0\rangle
        \label{eq:state}
\end{equation}
with $\{a_i\}$ an ordered set of integers between $1$ and $L$.
Note that now a particle may occupy the first site and a link
may be divided between sites $1$ and $L$.
These states will now be mapped onto the states of a reduced
chain with the number of sites being
\begin{equation}
         \tilde{L}=N_{\bullet \, \bullet}+N_{-  \, -}+N_{- \, \bullet}-
         N_{\downarrow \uparrow}
\end{equation}
where $N_{\downarrow \uparrow}$ is the total number of
$( \downarrow \uparrow)$
domain walls in the sequence of spins obtained by the mapping
above.
The above relation leads to the following relation between the
real and effective chain sizes
\begin{equation}
         \tilde{L}=L-N_{- \, \bullet}-N_{\downarrow \uparrow}
\end{equation}
which reflects the fact that our moving ``particles"  are now the
links $(- \, \bullet)$ with the exception of the ones which are domain
walls.
The two sites that compose such a link are effectively reduced to one
(or zero, if the link is a domain wall),
with the consequent reduction of the chain effective size.
Note that $\tilde{L}$ is always larger than $N_e$ or $N_h$.

The state given in Eq.~\ref{eq:state} corresponds to the following state
of the reduced chain
\begin{equation}
        \vert \tilde{a}_1, \dots ,
        \tilde{a}_{N_\downarrow+N_\uparrow};
        \sigma_1,\dots ,
        \sigma_{N_\downarrow+N_\uparrow}
        \rangle = \prod_{i=1}^{N_\downarrow+N_\uparrow}
        c_{\tilde{a}_i \sigma_i}^\dagger \vert \tilde{0}\rangle
\end{equation}
If site
$\tilde{1}$ is empty, the first link of the chain of $L$ sites
is  $(-  \, \bullet)$ and in order to have a well defined mapping, we impose
the condition that first two sites of the chain of $L$ sites correspond
to the link and therefore the first site is empty while the second is occupied.
The same applies in case of the site $\tilde{1}$ being occupied. Links
which are domain walls are not mapped to the reduced chain
(see Fig.~1 for an example of the mapping).
This condition agrees with the definition of states of the previous section
and furthermore, it also implies that certain states are not included in the
mapping, but, as previously, they can be written as translations of states
included in the mapping. These states which need to be translated appear
due to hoppings between sites $\tilde{1}$ and $\tilde{L}$, but also
sites $\tilde{1}$ and $\tilde{2}$.
As previously, we construct states invariant
by translation with total momentum  $P$,
\begin{equation}
        \vert \{\tilde{a}\},\{\sigma\}, P \rangle =
        \vert \{a\},P \rangle =
        {1 \over \sqrt{L}} \sum_{j=1}^{L}
        e^{iPj} \hat{T}^{j-1} \vert \{a\}\rangle.
\end{equation}
and keep the same mapping.  The states which need to be translated lead to
$e^{\pm iP}$ terms in the mapped Hamiltonian.
So that one does not need to be concerned with the reordering of operators in
the real chain, we will consider $N_e$ odd. The $N_e$ even case can be solved
with minor modifications of the procedure below.
Let $N_{\tilde{e}} = N_\downarrow+N_\uparrow$.

Note that in general, the hopping of an electron implies simply
that $\tilde{a}_j \rightarrow \tilde{a}_j \pm 1$ for some $j$.
Hoppings of a particle from 1 to 2 or 1 to L are however more
complex processes in the reduced chain. In the following tables,
we describe the action of these hopping terms. In the first column
of each table, one has the initial state and in the last column,
the final state after the application of the hopping operator. An
extra intermediate column is present if the final state can not be
directly mapped onto a state of the reduced chain. The second line
in each row shows the states in the original chain while the first
line shows the equivalent states in the reduced chain.

i) Let us first consider the jump of a link from $\tilde{1}$
to ${\tilde L}$. Note that this implies a $c^\dagger_L c_1$ hopping
for a link $(\bullet \, \bullet)$, but a $c^\dagger_1 c_L$
hopping for a link $(-  \, -)$.

\vspace{.5cm}
{\centering \begin{tabular}{|c|c|}
\hline
\( \uparrow \cdots \downarrow \circ  \)&
\( \circ  \cdots \downarrow \uparrow \)
\\
\hline
\( \bullet \bullet \cdots --\bullet - \)&
\( -\bullet \cdots --\bullet \bullet  \)
\\
\hline
\hline
\( \uparrow \cdots \uparrow \circ  \)&
\( \circ \cdots \uparrow \uparrow  \)
\\
\hline
\( \bullet \bullet \cdots \bullet \bullet - \)&
\( -\bullet \cdots \bullet \bullet \bullet  \)
\\
\hline
\hline
\( \uparrow \cdots \circ  \circ \)&
\( \circ \cdots \circ \uparrow  \)
\\
\hline
\( \bullet \bullet \cdots - \bullet - \bullet - \)&
\( -\bullet \cdots - \bullet - \bullet \bullet  \)
\\
\hline
\end{tabular}\par}
\vspace{.5cm}
{\centering \begin{tabular}{|c|c|c|}
\hline
\( \downarrow \cdots \uparrow \circ  \)&
no mapping &
\( e^{iP} \circ \cdots \uparrow \downarrow  \)\\
\hline
\( -- \cdots \bullet \bullet - \bullet \)&
\( \bullet - \cdots \bullet \bullet --  \)&
\(= T^{-1} - \bullet - \cdots \bullet \bullet - \)\\
\hline
\hline
\( \downarrow \cdots \downarrow \circ  \)&
no mapping &
\( e^{iP} \circ \cdots \downarrow \downarrow  \)\\
\hline
\( -- \cdots - - \bullet \)&
\( \bullet - \cdots ---  \)&
\(= T^{-1} - \bullet - \cdots - - \)\\
\hline
\hline
\( \downarrow \cdots \circ \circ  \)&
no mapping &
\( e^{iP} \circ \cdots \circ \downarrow  \)\\
\hline
\( -- \cdots - \bullet - \bullet \)&
\( \bullet - \cdots - \bullet --  \)&
\(= T^{-1} - \bullet - \cdots - \bullet - \)\\
\hline
\end{tabular}\par}
\vspace{.5cm}
ii) Now, the jump of a link from ${\tilde 2}$
to ${\tilde 1}$:

\vspace{.5cm}
{\centering \begin{tabular}{|c|c|}
\hline
\(\circ  \downarrow \cdots \downarrow \)&
\(\downarrow \circ  \cdots \downarrow \)
\\
\hline
\( - \bullet --\cdots - \)&
\( - - \bullet - \cdots -  \)
\\
\hline
\hline
\(\circ  \downarrow \cdots \uparrow \)&
\(\downarrow \circ \cdots \uparrow \)
\\
\hline
\( -\bullet -- \cdots  \bullet \bullet \)&
\( -- \bullet - \cdots  \bullet \bullet \)
\\
\hline
\hline
\(\circ  \downarrow \cdots \circ \)&
\(\downarrow \circ \cdots \circ \)
\\
\hline
\( -\bullet -- \cdots  - \bullet \)&
\( -- \bullet - \cdots  - \bullet \)
\\
\hline
\end{tabular}\par}
\vspace{.5cm}
{\centering \begin{tabular}{|c|c|c|}
\hline
\(\circ  \uparrow \cdots \uparrow   \)&
no mapping &
\( e^{iP} \uparrow  \circ \cdots  \uparrow \)\\
\hline
\( - \bullet \bullet \cdots \bullet  \bullet \)&
\( \bullet -\bullet \cdots \bullet \bullet  \)&
\(= T^{-1} \bullet \bullet -\bullet \cdots \bullet  \)\\
\hline
\hline
\(\circ  \uparrow \cdots \downarrow   \)&
no mapping &
\( e^{-iP} \uparrow \circ \cdots  \downarrow \)\\
\hline
\( - \bullet - \bullet \bullet \cdots - \)&
\( - \bullet \bullet -\bullet \cdots -  \)&
\(=  T^{+1} \bullet \bullet -\bullet \cdots -- \)\\
\hline
\end{tabular}\par}
\vspace{.5cm}
\noindent The last two cases also occur if the last pseudo-spin
is not at site $\tilde{L}$.
The Hamiltonian $H_1=H-V \cdot N_{\bullet \, \bullet}$,
in the mapped Hilbert space (in the subspace of momentum $P$), becomes
\begin{equation}
     H_1(P) = -  \sum_{\tilde{i}=1}^{\tilde{L}} t_{i\sigma}
     (1-n_{\tilde{i}\bar{\sigma}}) \tilde{c}_{\tilde{i}\sigma}^\dagger
     \tilde{c}_{\tilde{i}+1\sigma} (1-n_{\tilde{i}+1\bar{\sigma}})
      +hc
,
      \label{eq:H5}
\end{equation}
with
\begin{eqnarray}
      t_{\tilde{L}\uparrow} &=& t (-1)^{N_h}, \nonumber \\
      t_{\tilde{L}\downarrow} &=& te^{iP}(-1)^{N_h}, \nonumber \\
      t_{\tilde{1}\downarrow} &=& t, \nonumber \\
      \hat{t}_{\tilde{1}\uparrow} &=& 
      te^{\sigma_{N_{\tilde{e}}}\cdot iP}, \nonumber
\end{eqnarray}
and $t_{i\sigma}=t$, in the other cases.
This is the $U=\infty$ Hubbard chain pierced by a magnetic flux.
The Hamiltonian does not change the sequence
of spins $\{\sigma \}$, but circularly permutes them.
Note that $(-1)^{N_h}=(-1)^{\tilde{L}-N_\uparrow}$, if $N_e$ is odd.
In particular,
if $N_\uparrow=0$, this factor reflects the fact that a hole band
is translated by $\pi$ in relation to an electron band.

The solution of the above model is a little trickier than that of
the usual
$U=\infty$ Hubbard model\cite{dias1} due to the
term $\sigma_{N_{\tilde{e}}}$
in $\hat{t}_{\tilde{1}\uparrow}$.
Its solution is easier to understand if one considers first the
application of the Hamiltonian in the subspace of states with
the same configuration of the $\sigma$-spins, $\vert \sigma_1,\dots ,
\sigma_{N_{\tilde{e}}} \rangle $.
Then the Hamiltonian can be written in more compact notation, dropping
the spin index,
\begin{eqnarray}
     H_1(P) &=& -t  \sum_{\tilde{i}\neq \tilde{L}}
    \tilde{c}_{\tilde{i}+1}^\dagger
     \tilde{c}_{\tilde{i}}   \nonumber \\
     && \quad  -t_{\tilde{1}\sigma_1}
     \tilde{c}_{\tilde{1}}^\dagger
     \tilde{c}_{\tilde{2}}
     -t_{\tilde{L}\sigma_1}
     \tilde{c}_{\tilde{L}}^\dagger
     \tilde{c}_{\tilde{1}} \hat{Q}
      +hc.
      \label{eq:H6}
\end{eqnarray}
with hopping integrals given as above and $\hat{Q}$ being the cyclic 
spin permutation operator.

Consider a general state with no link at site $\tilde{1}$,
$   \vert \tilde{a}_1, \dots , \tilde{a}_{N_{\tilde{e}}}; \sigma_1,\dots ,
\sigma_{N_{\tilde{e}}} \rangle $.
If we redefine these states in the following way,
\begin{eqnarray}
      \vert \tilde{a}_1, \dots , \tilde{a}_{N_{\tilde{e}}}; \uparrow,\dots ,
      \sigma_{N_{\tilde{e}}} \rangle & \rightarrow & \\
      & & \hspace{-2cm} e^{\sigma_{N_{\tilde{e}}} iP}
      \vert \tilde{a}_1, \dots , \tilde{a}_{N_{\tilde{e}}}; \uparrow,\dots ,
      \sigma_{N_{\tilde{e}}} \rangle \nonumber
\end{eqnarray}
with $\tilde{a}_1 \geq 2$, the Hamiltonian within the subspace
of states with the above spin configurations becomes the one given
by Eq.~\ref{eq:H5} with the following modifications
\begin{eqnarray}
       t_{\tilde{1}\sigma} & \rightarrow & t,   \label{eq:ttilde} \\
        \hat{t}_{\tilde{L}\sigma} & \rightarrow & (-1)^{N_h} te^{1/2
       (1+\sigma_1 \cdot \sigma_{N_{\tilde{e}}})iP} .  \nonumber
\end{eqnarray}
The hoppings across the boundary do a cyclic permutation of the spin sequence
$\{\sigma \}$ with the above  phase factor.
We wish to construct now the states that remain invariant under such a
cyclic permutation, that is,
\begin{equation}
        \hat{Q}_{\{ \sigma \}}  \left( \sum_{i=0}^{r_{\alpha_c}-1} a_i
        \hat{Q}^i \vert \{\nu\}\rangle \right) = e^{i\phi'}
        \left( \sum_{i=0}^{r_{\alpha_c}-1}
        a_i \hat{Q}^i\vert \{\nu\}\rangle \right),
\end{equation}
where
\begin{eqnarray}
       \hat{Q}_{\{ \sigma \}} \vert  \sigma_{\tilde{1}} ,\dots ,
       \sigma_{N_{\tilde{e}}} \rangle  &= &\\
       && \hspace{-2cm}
       (-1)^{N_h} e^{1/2
       (1+\sigma_1 \cdot \sigma_{N_{\tilde{e}}})iP}
       \hat{Q} \vert  \sigma_{\tilde{1}} ,\dots ,
       \sigma_{N_{\tilde{e}}} \rangle \nonumber
\end{eqnarray}
and $r_{\alpha_c}$ is the periodicity of the spin configuration
and $\alpha_c$ labels the different spin configurations.
For example, the spin periodicity in
$\downarrow  \circ \downarrow \uparrow \downarrow  \downarrow  \circ \uparrow$
is 3.
$\phi'$ will be
the effective flux felt by the noninteracting fermions.
This problem is equivalent to solving a one-particle tight-binding model
for a chain of $r_{\alpha_c}$ sites with
hopping constant $t_{j}= t e^{1/2
(1+\sigma_1 \cdot \sigma_{N_{\tilde{e}}})iP}$, with the
correspondence $\vert i \rangle =\hat{Q}^{i-1} \vert \{\nu \} \rangle$.
The total flux through this tight-binding chain is
\begin{equation}
        \phi_1= r_{\alpha_c} {N_\uparrow+N_\downarrow 
        -2N_{\downarrow \uparrow}
        \over N_\uparrow+N_\downarrow} iP .
\end{equation}
The solution is obtained after a gauge transformation so that
$t_{j} \rightarrow e^{i\phi_1 /r_{\alpha_c}} t$. The gauge transformation
depends on the $\nu$-spin configuration, but the tight binding eigenvalues
only depend on the total flux. The eigenstates will be Bloch states $\vert
\alpha_c, q_c \rangle$ (in the cyclic permutations)
with $q_c=n(2\pi/r_{\alpha_c})$, with $n=0,\dots,r_{\alpha_c}-1$.
This resolution is rather similar to that of the Hubbard model with flux
which has been treated in Ref.~\cite{nuno1}.

Its solution is known \cite{dias1,nuno1}
and the eigenvalues of $H_1$ for $L$ odd are given by
\begin{equation}
        E(\{\tilde{k}\},q_c,P)=-2t \sum_{i=1}^{N_{\tilde e}}
        \cos \left(\tilde{k}_i+
        \alpha {P \over \tilde{L}}+{q_c  \over \tilde{L}}\right)
        \label{eq:eigenvalues}
\end{equation}
with
\begin{equation}
        \alpha = {N_\uparrow+N_\downarrow -2N_{\downarrow \uparrow}
        \over N_\uparrow+N_\downarrow}.
\end{equation}
If $L$ is even, there is a $\pi/\tilde{L}$ correction in the argument of the
cosine due to the term $(-1)^{N_h}$. Note the sign change within the cosine
argument when compared with Eq.~\ref{eq:sevE}.
This sign change just reflects the ``particle-hole"
transformation which is implicit in the fact that now the links $(- \bullet)$
are mapped onto holes.

Now, the total momentum $P$ has to be determined as a
function of $\{ k \}$ and $q_c$.
The following condition
is obtained from the phase acquired by a eigenstate
under the translation of two real sites or a pseudo-site,
\begin{equation}
         2P =\sum_{i=1}^{N_e}(\tilde{k}_i + \alpha
        {P \over \tilde{L}}+{q_c \over \tilde{L}})
        \quad (\text{mod} \,\, 2\pi)
\end{equation}
which is easy to understand  examining the translation
of a component of the eigenstate
which does not have pseudo-particles at site $\tilde{L}$
and therefore does not suffer a circular permutation of the
pseudo-spins.
Obviously, the components that do not satisfy the previous
assumption will lead to the same result since the overall
eigenstate is invariant by translation.
This relation  can be written in a simpler form
\begin{equation}
        P {L \over \tilde{L}}=\sum_{i=1}^{N_e} \tilde{k}
        + (N_\uparrow+N_\downarrow) {q_c \over \tilde{L}}
        \quad (\text{mod} \,\, 2\pi).
        \label{eq:mom}
\end{equation}
As in the previous section, the set of pseudo-momenta $\{\tilde{k}\}$
and the pseudo-spin momentum $q_c$ are not enough to define totally $P$.

\begin{figure}[htb] 
       \begin{center} 
       \leavevmode 
       \hbox{%
       \epsfxsize 2.5in \epsfbox{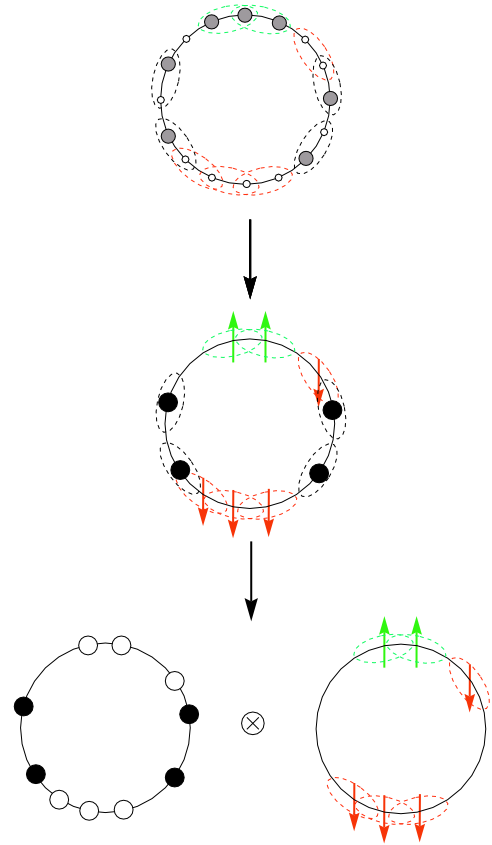}} 
       \end{center} 
       \caption{The mapping is illustrated in this Figure. 
       In the first chain, the circles stand for occupied sites 
       while the small dots stand for empty ones. 
       The introduction of the nearest-neighbor interaction 
       leads to a further factorization of 
       the wave function describing the charge 
       degrees of freedom.} 
       \label{fig:factor} 
\end{figure}

The spin-charge factorization of the $U=\infty$
Hubbard model translates into a decoupling of the degrees
of freedom describing the configuration of domains
of
solitons and antisolitons
and the degrees of freedom associated to the presence of ``holes''
moving through these domains.
This factorization and the mapping
presented in this paper are illustrated in Fig.~\ref{fig:factor}.

\subsection{Flux dependence}

Assume now that the chain is pierced by an 
external flux $\phi$, that is,
the Hamiltonian is given by Eq.~\ref{eq:H2} 
with $t \rightarrow t e^{i \phi /L}$.
This problem can be solved following the same procedure as for $\phi=0$
with an extra step. This step is equivalent to the gauge transformation
$$
       {\bar c}_{j }^\dagger = c_{j }^\dagger e^{i \phi /L \cdot j}
$$
which carries all the phase to hoppings at the boundary,
$t_{j} \rightarrow t; \quad j \neq L$, $t_{L} \rightarrow t e^{i \phi}$.
Let us show how this can be done for the mapped Hamiltonian.
We modify the state invariant by translation in the following way,
\begin{eqnarray}
        \vert a_1, \dots , a_{N_e},P \rangle &=& \\
        && \hspace{-2cm} e^{i {\phi \over L} \sum_{i=i}^{N_e} a_i}
        {1 \over \sqrt{L}} \sum_{j=1}^{L}
        e^{iPj} \hat{T}^{j-1} \left(
        \prod_{i=i}^{N_e} c_{a_i }^\dagger \vert 0\rangle \right) . \nonumber
\end{eqnarray}
Now note the following,
\begin{eqnarray}
        ( \bullet - \cdots \bullet \bullet -- )_P &=&
        e^{iP} e^{-iN_e{\phi \over L}} \quad
        (- \bullet - \cdots \bullet \bullet - )_P ;
        \nonumber  \\
        ( \bullet - \bullet \cdots \bullet \bullet )_P \quad &=&
        e^{iP} e^{-iN_e{\phi \over L}} e^{i\phi}
        ( \bullet \bullet - \bullet \cdots \bullet )_P.
        \nonumber
\end{eqnarray}
Therefore, we will have an extra phase term in the hoppings displayed
in the previous tables which involve a translation. Furthermore, a hopping of
a link $(\bullet \bullet)$ at the boundary implies a hopping of an electron
in the same direction while the hopping of  a  link $(--)$ implies
a hopping of an electron  in the opposite direction. For zero external
flux, this distinction
would be irrelevant, but for a finite flux, it leads to a spin dependent
phase of the hopping integral $e^{-\sigma_1 i \phi}$.
Following exactly the same procedure, we arrive to the same stage of
Eq.~\ref{eq:H5} with the following modifications
\begin{eqnarray}
         t_{\tilde{L}\uparrow}&\rightarrow & t_{\tilde{L}\uparrow}
         e^{-i \phi}
;
         \nonumber  \\
         t_{\tilde{L}\downarrow}&\rightarrow & t_{\tilde{L}\downarrow}
         e^{i \phi}  e^{- iN_e{\phi \over L}} ; \nonumber \\
         t_{\tilde{1}\downarrow}&\rightarrow & t_{\tilde{1}\downarrow} ;
         \nonumber  \\
         t_{\tilde{1}\uparrow}&\rightarrow & t_{\tilde{1}\uparrow}
         e^{- \sigma_{N_{\tilde e}} \cdot iN_e{\phi \over L}}
         e^{ (1+\sigma_{N_{\tilde e}}) \cdot i\phi /2 } . \nonumber
\end{eqnarray}
Following the same steps, this leads to the modification
$$
       t_{\tilde{L}\sigma} \rightarrow t_{\tilde{L}\sigma}
       e^{-\sigma_1 i \phi} e^{-1/2
       \cdot
       (1+\sigma_1 \cdot \sigma_{N_{\tilde{e}}})iN_e{\phi \over L}}
       e^{1/4 \cdot (1+\sigma_{N_{\tilde e}})(1+\sigma_1)\cdot i\phi}.
$$
This phase term generates an extra flux contribution through the
$(N_\uparrow+N_\downarrow)$ tight-binding chain which is given by
\begin{equation}
    \left(
    -{N_\uparrow-N_\downarrow \over N_\uparrow+N_\downarrow}
    -\alpha {N_e \over {\tilde L}} +
    {N_{\uparrow \uparrow} \over N_\uparrow+N_\downarrow}
    \right) \cdot \phi
\end{equation}
where $N_{\sigma \sigma'}$ is the number of pairs $\sigma \sigma'$
in the sequence
of spins obtained with our mapping. For example, in Fig.~1,
$N_{\uparrow\uparrow}=1$, $N_{\downarrow\downarrow}=3$ and
$N_{\uparrow\downarrow}=1$. Note that
$N_\uparrow = N_{\uparrow\uparrow}+ N_{\uparrow\downarrow}$,
$N_\downarrow =  N_{\downarrow\downarrow } + N_{\uparrow\downarrow}$,
$N_{\uparrow\downarrow} = N_{\downarrow\uparrow}$.
The Hamiltonian is simple to diagonalize and the eigenvalues are given
by Eq.~\ref{eq:eigenvalues} with
\begin{equation}
        E(\{\tilde{k}\}) \rightarrow E \left( \left\{\tilde{k}-
        \beta {\phi \over L} \right\} \right)
        \label{eq:flux}
\end{equation}
and
\begin{eqnarray}
       \beta&=&{N_{\uparrow\uparrow}{N_e \over \tilde{L}}-
       N_{\downarrow\downarrow }{N_h \over \tilde{L}}
       \over N_\uparrow+N_\downarrow}
       \nonumber \\
       &=&  {N_\uparrow-N_\downarrow \over N_\uparrow+N_\downarrow}
.
\end{eqnarray}
Such expression for the flux dependence 
should be expected since solitons and anti-solitons 
in the strong coupling limit can be viewed as hard-core particles 
with opposite charges and a simple spinless model of hard-core 
particles with opposite charges in a magnetic flux would exhibit 
precisely this flux dependence of the eigenvalues.
It is easy to show that if $N_{\uparrow}=0$ or  $N_{\downarrow}=0$,
$\beta=\pm 1$ as it should be. One can interpret $\beta$ as the 
effective charge
of the carriers.
Note that this renormalization of the flux dependence was also
found for the strong coupling Hubbard model\cite{nuno1} with
precisely the same form.

\subsection{Higher order corrections}
The second order corrections can be obtained considering
virtual hoppings
that  create or destroy a soliton-antisoliton pair.
For a given low lying eigenstate with $N_\uparrow=0$, this leads
to a energy correction of the form
\begin{equation}
       {t^2 \over V} \langle n_{i\downarrow} n_{i+1 \downarrow}
       +(1-n_{i+1 \downarrow})
       (c^\dagger_{i\downarrow} c_{i+2 \downarrow} +
       c^\dagger_{i+2 \downarrow} c_{i\downarrow}) \rangle.
\end{equation}
When $N_\downarrow=0$, the energy correction is of the
same form. If $N_\uparrow+N_\downarrow=\tilde{L}$, the
second-order corrections can be mapped on a Heisenberg spin model.
In the general case, the energy correction can be written as an
average over an operator that creates (or destroys) a
soliton-antisoliton pair and destroys (or creates) also a
pair which may or may not be the one created (destroyed), leading
to long range hopping of these pairs with or without exchange of
the pair. A closer mapping than that onto the $U=\infty$ Hubbard
model is suggested at this level, since the  above corrections are
also present in the charge sector of the $U \gg t$ Hubbard model,
if the spin configuration is restricted to be Neel like with 
momentum $q_s=0$. In this case, long range hopping of a 
hole-``double occupancy'' pair is also possible and one may 
think of doubly occupied sites, holes and singly occupied sites
(with an $q_s=0$ Neel configuration) as equivalent to $(\bullet \,
\bullet)$, $(-  \, -)$ and empty sites in our reduced chain. The
flux dependence of the eigenvalues also suggests such a
picture.\cite{nuno1} We will see that such a picture agrees with
the transport properties of the t-V model.
\section{Comparison with Bethe ansatz results}
Our results can be linked to those obtained with the Bethe ansatz
technique. \cite{nuno2,kusm}  In the following, we adopt the
notation of Ref.~\cite{nuno2}. The Bethe ansatz solution is
characterized by a set of bands $\gamma$  (with
$\gamma=0,1,\dots$), with non-trivial relations for the total
number of available ``single-particle'' states in each band
$d_{c,\gamma}$ and for the total number of occupied states in each
band, $N_{c,\gamma}$. The energy associated with an occupied state
in $\gamma \neq 0$ band is of order $V$ and therefore, the
$\gamma=0$ band is the free carrier band obtained in our
picture. The high energy bands ($\gamma\neq 0$) are related to the
remaining degrees of freedom associated with the possible
configurations of the links $(\bullet \, \bullet)$ and  $(-  \,
-)$. This is similar to the strong coupling Hubbard model case
where the high energy BA bands are clearly linked to the
possible
configurations of holes and double occupancies.\cite{nuno1}

In Table 1 of Ref.~\cite{nuno2}, we see that the low
lying states ($N_0 \neq 0$,
$N_\gamma=0$, for $\gamma>0$) are those
of a chain  with a reduced size ${\tilde L}=d_{c,0}=L-N_e$ 
and number of holes
given by $N_{\tilde h}=N_{c,0}^h=L-N_e$,
which agrees with our equivalent findings
in Section \ref{sec:particles}.
Noting that in Eq.~24 of Ref.~\cite{nuno2},
$$
    {2 \pi \over L N_e}-{2 \pi \over (L-N_e)N_e}
    =- {2 \pi \over L (L-N_e)}
,
$$
the eigenvalue expression, Eq.~23 of Ref.~\cite{nuno2},
becomes exactly the same as our Eq.~\ref{eq:sevE}
and the same can be said for the flux
dependence of these eigenvalues.

The high energy states are more complex since they are
characterized  by a non-zero occupation of the high energy bands.
Let us assume for simplicity that only one of the high energy BA
bands ($\gamma\neq 0$) is occupied. The effective size of the
chain and the number of holes in the $\gamma= 0$ band are\cite{nuno2}
\begin{eqnarray}
    {\tilde L}&=& d_{c,0} = L-N_e+(\gamma-1)N_{c,\gamma}, \nonumber \\
    N_{\tilde e}&=& N_{c,0}^h = L-2N_e+2\gamma N_{c,\gamma}
,
\end{eqnarray}
where in the last equation, we have used the fact the holes in the $\gamma= 0$
band are in our picture the particles.
Relating these two equations to the definition 
of these quantities in our picture,
one obtains
\begin{equation}
    \gamma N_{c,\gamma}= N_{\bullet \, \bullet}; \quad
    N_{c,\gamma}= N_{\uparrow \downarrow}
.
\end{equation}
These relations can be confirmed calculating the total number of electrons
\begin{eqnarray}
    N_e&=& N_{c,0}+ (\gamma+1)N_{c,\gamma}\nonumber \\
    &=&N_{- \, \bullet}-N_{\uparrow \downarrow}
    +N_{\bullet \, \bullet}+N_{\uparrow \downarrow}
.
\end{eqnarray}
These relations imply that $\gamma$ has a very simple physical meaning
in the strong coupling limit,
it is the size of the clusters of links $(\bullet \, \bullet)$.
Since only one BA band is occupied, all clusters have the same size
and the total number of these clusters is $N_{\uparrow \downarrow}$.
The total number of links $(\bullet \, \bullet)$ is then obviously
$\gamma \cdot N_{\uparrow \downarrow}$.
This type of excitations form the so-called BA strings,\cite{string1,string2}
and in particular, the string associated with an occupied state in band
$\gamma$ has length $\gamma$ (see Ref.\cite{string1} for an explanation
of these BA string excitations and of the precise meaning of string length).
We see now that, for t-V model in the strong
coupling limit, a string is simply a cluster
of links $(\bullet \, \bullet)$ in the configuration of
links $(\bullet \, \bullet)$ and $(- \, -)$.
In the general case, several BA bands are occupied implying a configuration
of links $(\bullet \, \bullet)$ and $(- \, -)$ where the links
$(\bullet \, \bullet)$ combine into clusters of several lengths.
\section{Transport properties}
The transport properties of one-dimensional models have acquired a renewed
interest recently due to a conjecture by Zotos 
{\it et al}\cite{zotos,cast,naef}
that integrable
models with zero Drude weight at zero temperature are ideal insulators,
that is, the Drude weight remains zero also at finite temperature.
Based on qualitative arguments, Zotos and Prelovsek\cite{zotos}
have also stated
that, in the particular case of the strong coupling half-filled
t-V model, such temperature independence
is present even for finite size chains.
This has been confirmed by a Bethe ansatz study of this model.\cite{nuno2} 
These
results can be easily rederived with our solution and they are simple
consequences of the $\beta$ prefactor in the flux dependence of
Eq.~\ref{eq:flux}.
For instance, the current operator
\begin{equation}\label{current}
      J=it\sum_{i} (c_{i}^\dagger c_{i+1}
      - c_{i+1}^\dagger c_{i})
\end{equation}
can  be obtained at finite temperatures from
\begin{equation}
      \langle j \rangle =- {1 \over 2} \sum_n {e^{- \beta E_n} \over Z}
      {\partial(E_n/L) \over \partial(\phi_c/L)} \mid_{\phi=0}
\end{equation}
and therefore, if all eigenvalues are flux independent, the current
will be zero whatever the temperature value.
Note this is a stronger absence of current than the usual
situation, which may occur also in metallic systems, 
where the zero average of the current operator results from the fact that 
the positive energy slopes being exactly compensated by the negative ones.
Also, the charge stiffness is given by \cite{khon,shas}
\begin{equation}
       D_c={1 \over 2} \sum_n {e^{- \beta E_n} \over Z}
       {\partial^2(E_n/L) \over \partial(\phi_c/L)^2} \mid_{\phi_c=0}.
\end{equation}
and as for the current, if all eigenvalues are flux independent,
the Drude weight remains zero at finite temperatures.
A eigenvalue in order to be flux independent must have
$N_{\uparrow}- N_{\downarrow}=0 $.
It is easy to show that is indeed the case for half-filled states.
For these states, $N_e=N_h$ and since $N_{\uparrow}-
N_{\downarrow}= N_e-N_h$,
$\beta=0$.

In the following, we illustrate the use of our solution
with a study of the optical conductivity.
The real part of the conductivity $\sigma (\omega)$ is given by
\begin{equation}
        \sigma(\omega)=2\pi D_c
        \delta(\omega) +\sigma_{reg}(\omega)
\end{equation}
with
\begin{eqnarray}\label{reg}
        \sigma_{reg}(\omega) = {1-e^{-\beta \omega} \over \omega}
        {\pi \over L} \sum_{n,m \neq n}
        && p_n
        \vert \langle n\vert J \vert m\rangle \vert^2 \nonumber \\
        &&\cdot  \delta (\omega -E_m +E_n)
\end{eqnarray}
where $p_n$ is the Boltzmann weight.

At half-filling, the ground state of the t-V model
is insulating ($D_c(0)=0$) and doubly
degenerate, one state having momentum $0$, the other $\pi$.
Both states have $N_{- \bullet}=L/2$ ($L$ even) and
$N_{\bullet \bullet}=N_{--}=0$.
The current operator applied to a ground state
induces transitions to states with
$N_{\bullet \bullet}=N_{--}=1$ and $N_{- \bullet}=L/2-2$.

One should note that when determining the optical conductivity
at finite temperature,
one has to calculate matrix elements of the current operator
between states with $N_{\bullet \bullet}$ and
$N_{\bullet \bullet}'=N_{\bullet \bullet}+1$
in order to obtain the upper band part of the optical
conductivity.
The low frequency region is given by matrix elements of the current
operator between states with the same number of links.
Clearly, the contribution of the states $\vert n\rangle
=\vert N_{\bullet \bullet}\neq 0\rangle$ will
be very small as its Boltzmann weight
is $p_n \sim e^{-\beta N_{\bullet \bullet}U}/Z$
and we will only consider temperatures
$T \ll V/k_B$.
So, the sum over $\vert n\rangle$ becomes a sum over all states
$\vert N_{\bullet \bullet}= 0\rangle$ and
the sum over $\vert m\rangle$ becomes
a sum over all states $\vert N_{\bullet \bullet}= 0\rangle$
for the low frequency conductivity
and a sum over $\vert N_{\bullet \bullet}= 1\rangle$
for the upper band part of the
conductivity.
That is, we can write $\sigma_{reg}=\sigma_o+\sigma_1$, where $\sigma_o$
will be the low frequency conductivity ($\omega \sim t$)
\begin{eqnarray}
        \sigma_o(\omega)&=& {1 \over \omega}
        {\pi \over L} \sum_{N'_{\bullet \bullet} = 0 \atop
        N_{\bullet \bullet} = 0} p_{ N_{\bullet \bullet} = 0}  \\
        &&\vert \langle N'_{\bullet \bullet} = 0\vert J_0 \vert
        N_{\bullet \bullet} = 0\rangle \vert^2
        \delta (\omega -E_{N'_{\bullet \bullet} = 0}
        +E_{N_{\bullet \bullet} = 0})
        \nonumber
\end{eqnarray}
and  where $\sigma_1$
will be the high frequency conductivity ($\omega \sim V$)
\begin{eqnarray}
        \sigma_1(\omega)&=& {1-e^{-\beta V} \over V}
        {\pi \over L} \sum_{N'_{\bullet \bullet} = 1 \atop
        N_{\bullet \bullet} = 0} p_{N_{\bullet \bullet} = 0}  \\
        &&\vert \langle N'_{\bullet \bullet} = 1\vert J_1 \vert
        N_{\bullet \bullet} = 0\rangle \vert^2
        \delta (\omega -E_{N'_{\bullet \bullet} = 1}
        +E_{N_{\bullet \bullet} = 0})
        \nonumber
\end{eqnarray}
where $J=J_0+J_1$, $J_0$ being the part of the current operator
which does not alter the number of links and therefore,
commutes with the strong coupling Hamiltonian,
\begin{eqnarray}
     J_0 &=& it \sum_{i} (1-n_{i+2}) c_{i+1}^\dagger c_{i} (1-n_{i-1})
    \nonumber \\
     &+& it \sum_{i} n_{i+2} c_{i+1}^\dagger c_{i} n_{i-1}  +hc,
\end{eqnarray}
and $J_1$ being the sum of terms in the current operator which
induce transitions between states such that their energies differ by $V$,
\begin{eqnarray}
     J_1 &=& it \sum_{i} (1-n_{i+2}) c_{i+1}^\dagger c_{i} n_{i-1}
    \nonumber \\
     &+& it \sum_{i} n_{i+2} c_{i+1}^\dagger c_{i} (1-n_{i-1})  +hc.
\end{eqnarray}
Since $J_0$ commutes with the Hamiltonian,
$\sigma_o(\omega)= 0$.

In this paper, we will only evaluate $\sigma_1(\omega)$ at zero
temperature and half-filling ($D_c=0$),
\begin{eqnarray}
        \sigma_1(\omega) = {1 \over V}
        {\pi \over L} \sum_{N'_{\bullet \bullet} = 1} && \nonumber \\
        && \hspace{-2cm}
        \left(
        \left\vert \left\langle \left.
        {N'_{\bullet \bullet} = 1 ; P=0}
        \right\vert
        J_1
        \left\vert
        {N _{- \bullet}=N_e ; P=0 }
        \right. \right\rangle \right\vert^2
        \right.
        \nonumber \\
        && \hspace{-2cm} +
        \left.
        \left\vert \left\langle \left.
        {N'_{\bullet \bullet} = 1 ; P= \pi }
        \right\vert
        J_1
        \left\vert
        {N _{- \bullet}=N_e ; P= \pi }
        \right. \right\rangle \right\vert^2 \right)\nonumber \\
        && \hspace{-2cm} \cdot \,
        \delta (\omega -E_{N'_{\bullet \bullet} = 1} )
        \nonumber
\end{eqnarray}
where $\vert N_{- \bullet}=N_e; P= \pi  \rangle $ and
$\vert N_{- \bullet}=N_e; P=0  \rangle $ are the two possible ground states
at half-filling.
Well,
\begin{eqnarray}
     J_1 \vert N_{- \bullet}=N_e P=0  \rangle
     &=& it\sqrt{L} \vert 1,2; q_c(\downarrow \uparrow)=\pi;P=0  \rangle ,
    \nonumber \\
     J_1 \vert N_{- \bullet}=N_e P=\pi  \rangle
     &=&  it\sqrt{L} \vert 1,2; q_c(\downarrow \uparrow)=0;P=\pi \rangle ,
     \nonumber
\end{eqnarray}
so we only need to calculate the overlap
of the eigenstates with the above states.
Note that for these eigenstates, $N_\downarrow=N_\uparrow=1$ and therefore
$N_{\downarrow\uparrow}=1$. Consequently, $\alpha=0$.
The respective eigenvalues are given by
\begin{eqnarray}
      E(\tilde{k}_1,\tilde{k}_2)
      &=& V -2t \sum_{i=1}^2 \cos
      \left(\tilde{k}_i+{q_c-\pi \over \tilde{L}}\right)
       \\
      &=& V -4t  \cos \left({\tilde{k}_1+\tilde{k}_2 \over 2}
      +{q_c-\pi \over \tilde{L}} \right)
      \cos \left({\tilde{k}_1-\tilde{k}_2 \over 2} \right) .\nonumber
\end{eqnarray}
These eigenstates are given by
\begin{eqnarray}
      \vert \tilde{k}_1,\tilde{k}_2;q_c;P\rangle
      = {1 \over L}
      \sum_{j<l} &&
      \left( e^{i(\tilde{k}_1 j+\tilde{k}_2 l)}
      -e^{i(\tilde{k}_1 j+\tilde{k}_2 l)} \right) \nonumber \\
      && \cdot  e^{i {q_c -\pi \over L} (j+l)} \vert j,l;q_c;P \rangle .
\end{eqnarray}
From Eq.~\ref{eq:mom} and noting that $2P=0$ (mod $2\pi$), one has
$k_1+k_2=2\pi /\tilde{L}$ (mod $2\pi$) if $q_c=0$
and $k_1+k_2=0$ (mod $2\pi$) if $q_c=\pi$.
Well, $\sigma_1$ has two contributions of the form
\begin{equation}
      \sin^2 \left({\tilde{k}_1-\tilde{k}_2 \over 2} \right)
      \cdot \delta[\omega-E(\tilde{k}_1,\tilde{k}_2)]
,
\end{equation}
which taking into account the above conditions can be written as
$\sin^2 (\tilde{k}) \cdot \delta[\omega-V+4t \cos(\tilde{k})]$ in
both cases. The ground state momentum ($P=0$ or $P=\pi$) is
irrelevant as expected. So,
\begin{eqnarray}
      \sigma_1 &\sim & \sum_{\tilde{k}} \sin^2 (\tilde{k})
      \cdot \delta[\omega-V+4t \cos(\tilde{k})] \nonumber \\
      &\sim & \left[ 1- \left({\omega-V \over 4t} \right)^2 \right]
      \cdot N^{1d} \left({\omega-V \over 2} \right) 
\end{eqnarray}
where $N^{1d}$ is the density of states of a one-dimensional
tight-binding model. This density of states is non-zero between
$-2t$ and $2t$ and has inverse square root divergences at  $\pm
2t$. Therefore, the optical conductivity will be characterized by absence
of weight between zero and $V-4t$, which is the optical gap. At
the extremes, the optical conductivity goes to zero as $\sigma_1
\sim \sqrt{\vert \omega-V \vert -4t} $, that is,
\begin{eqnarray} 
      \sigma_1 &\sim & \sqrt{ 1- \left({\omega-V \over 4t} \right)^2 }, 
      \quad -4t<\omega-V<4t.
\end{eqnarray} 
This is precisely the
dependence obtained by Lyo and Galinar\cite{lyo1,gebh} for the
strong coupling Hubbard model with a Neel ground
state.\cite{lyo1,lyo2} Again, the t-V model seems to behave as the
strong coupling Hubbard model with a fixed Neel spin
configuration.
The optical conductivity of the strong coupling t-V model
has also been recently studied with a conjunction of Bethe ansatz and 
conformal invariance,\cite{carm} which has allowed 
the determination of the exponent for the   
frequency dependence immediately above the absorption edge.
The exponent obtained was $1/2$, in agreement with our results.
\section{Conclusion}
In conclusion, we have presented a non-Bethe-ansatz solution for
the strong coupling t-V model with twisted boundary conditions
(or equivalently, the strongly anisotropic Heisenberg
model).
We have found that this model can be described in terms of
soliton-antisoliton
configurations and non-interacting particles moving in
a reduced chain threaded by a fictitious flux generated by the previous
configuration, but also containing a term proportional to the total momentum
of the non-interacting particles.
The flux dependence of the eigenvalues  remains
unchanged for the low lying states, but is reduced for intermediate
energies reflecting the renormalization of the
charge of the non-interacting particles.
Much of the previous picture was obtained with a simple mapping
of this model onto the $U=\infty$ Hubbard model.
However, the t-V model remains simpler than the $U=\infty$ Hubbard model,
since no large spin degeneracy is present in the low
energy sector. This allows a much easier calculation of
correlations. As an example, we presented the simple calculation
of the zero temperature optical conductivity of this model 
at half-filling.

The arguments by Zotos and Prelovsek\cite{zotos} and the
Bethe ansatz studies of the t-V model\cite{nuno2} in the strong coupling limit
have been confirmed here. All states obtained from the half-filled
insulating ground-state by successive applications of the single
particle hopping operator have energies independent of the
external magnetic flux and consequently, the Drude weigth remains
zero even at finite temperature. 
However, one should note that the results presented here do not exclude 
a positive charge stiffness at finite temperature 
resulting from the $t^2/V$ corrections to the strong coupling Hamiltonian. 
Obviously, the magnitude of the charge stiffness (if non-zero) 
resulting from these corrections will be at the most of the order of $t^2/V$.
The flux independence is closely linked
with the soliton-antisoliton$\otimes$``free particles''
factorization at intermediate energies. It is curious to note that
such a factorization in the charge sector for intermediate
energies is also present in the strong coupling Hubbard model.\cite{nuno1} 
The
extended Hubbard model can also be solved in the strong coupling
limit taking a similar path to that presented here.\cite{dias3}

Finally, note that we have assumed implicitly throughout the paper that $V$
was positive, but, obviously, the solution is valid for both $V/t
\rightarrow \infty$ and $V/t \rightarrow -\infty$.
For $t=0$, the eigenstates of the model are the same, independently 
of the sign of the nearest-neighbor interaction, but obviously the 
respective eigenvalues are symmetric for $V$ positive or negative. 
For $t\neq 0$, the projection of the kinetic energy operator in the degenerate 
subspaces is independent of the sign of the interaction and so will be the 
diagonalization of this operator in each degenerate subspace. Therefore, 
the eigenstates are the same for $V/t \rightarrow \pm \infty$ 
with the respective eigenvalues being given by Eq. 42 plus or minus 
$N_{\bullet \bullet} \cdot \vert V \vert $, 
according to the sign of the interaction $V$. 
In particular, at half-filling, the phase separated state 
(which is the highest energy state when 
$V$ is positive) and the two charge ordered groundstates 
trade places when $V$ is negative, i.e., the phase 
separated state becomes the ground state and the two charge ordered 
states become the highest energy states.
\acknowledgements
We wish to
thank Nuno M. R. Peres and Joao M. Lopes dos Santos for important
discussions.
This research was funded by the Portuguese \linebreak
MCT PRAXIS XXI program under Grant No. 2/2.1/Fis/302/94.


\end{document}